\documentclass[preprint]{aastex}

\def\ltsima{$\; \buildrel < \over \sim \;$}
\def\simlt{\lower.5ex\hbox{\ltsima}}
\def\hb{{\sc{H}}$\beta$\/}

\received{} \accepted{}
\def\l{$\lambda$}

\def\msol{M$_\odot$\/}

\def\ltsima{$\; \buildrel < \over \sim \;$}
\def\simlt{\lower.5ex\hbox{\ltsima}}            
\def\gtsima{$\; \buildrel > \over \sim \;$}

\def\simgt{\lower.5ex\hbox{\gtsima}}            
 
\def\a{$\alpha$}

\def\hi{{\sc Hi}\/}
\def\hii{{\sc Hii}\/}
\def\ha{{\sc H}$\alpha$}

\def\dvr{$\Delta \rm v_r$\/}
\def\vr{$\rm v_r$\/}
\def\cm3{cm$^{-3}$\/}
\def\hb{{\sc{H}}$\beta$\/}

\def\o4363{{\sc{[Oiii]}}$\lambda$4363\/}

\def\fe{{\sc{Fe}}\/}

\def\fe76087{{\sc [Fe vii]}$\lambda$6087\/}

\def\nii{{\sc [Nii]}$\lambda\lambda$6548,6583}
\def\sii{{\sc [Sii]}$\lambda\lambda$6716,6731}
\def\b{$\beta$}
\def\kms{km~s$^{-1}$}
\def\ergss{ergs s$^{-1}$\/}
\def\vr{$v{\rm_r}$\/}
\begin{document}
\title{\sc Arp 194: Evidence of Tidal Stripping of Gas and
Cross-Fueling}

\author{P. Marziani\altaffilmark{1},  D. Dultzin-Hacyan\altaffilmark{2},
M. D'Onofrio\altaffilmark{3}, J. W. Sulentic\altaffilmark{4}}

\altaffiltext{1}{Osservatorio Astronomico di Padova, INAF, vicolo
dell'Osservatorio 5, I-35122 Padova, Italy; marziani@pd.astro.it}

\altaffiltext{2}{Instituto de Astronom\'\i a, UNAM, M\'exico, D.
F. 04510, M\'exico; deborah@astroscu.unam.mx}

\altaffiltext{3}{Dipartimento di Astronomia, Universit\`a  di
Padova, Vicolo dell'Osservatorio 3, I-35122 Padova, Italy;
donofrio@pd.astro.it}

\altaffiltext{4}{Department of Physics and Astronomy, University of
   Alabama, Tuscaloosa, AL 35487, USA; giacomo@merlot.astr.ua.edu}

\begin{abstract}  We present new imaging and spectroscopic observations of the
interacting  system Arp 194 ($\equiv$ UGC 06945  $\equiv$ VV 126).  The
northern component (A194N) is a distorted spiral or ring galaxy likely
disrupted by a collision or close encounter with a southern galaxy (A194S).
There is evidence that a third galaxy with similar recession velocity is
projected on A194N but its role is likely secondary. A194S is connected to
A194N by a string of emission knots which motivates our interpretation that
the former was the intruder. Three of the knots are easily discernible in
B,R, and \ha\ images and are assumed to trace the path of the intruder
following the encounter, which we estimate occurred a few 10$^8$ yr ago.

Both A194S and N are experiencing strong bursts of star formation: the \ha\
luminosity indicates a total star formation rate $\sim$ 10 \msol
yr$^{-1}$.  The lack of detectable J and K emission from the blobs, along
with strong \ha\ emission, indicates that an evolved stellar population is
not likely to be present. The brightest knot (closest to A194S) shows a star
formation rate of $\approx$1.2 \msol yr$^{-1}$\ which, if sustained over a
time $\approx$ 7$\times 10^7$ yr, could explain the spectral energy
distribution. This suggests that the stripped matter was originally
predominantly  gaseous. The brightest knot is detected as a FIRST radio
source and this is likely the signature  of supernova remnants  related to
enhanced star formation.  Motions in the gas between the brightest knot and
A194S, traced by an emission line link of increasing radial velocity,
suggests infall toward the center of the intruder.  Arp 194 is therefore
one of the few galaxies where evidence of ``cross-fueling'' is observed.

\end{abstract}


\section{Introduction}

\begin{figure}
\plotone{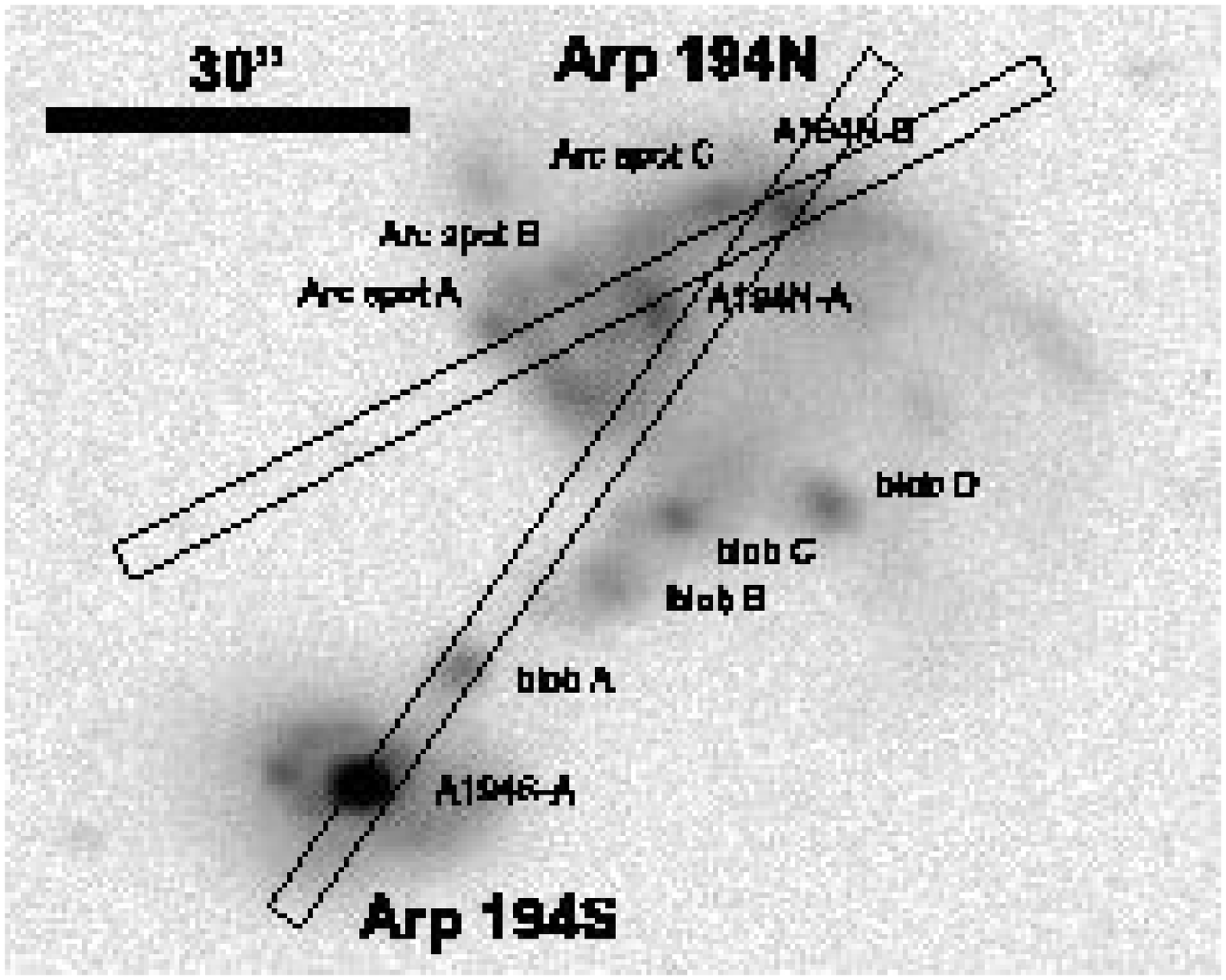} \figcaption[ident01.eps]{Morphological details
of the Arp 194 system and their identification. The slit position (P.A.
=145$^\circ$\ and P.A. =118$^\circ$) and width (2.6") is shown on scale.
North is to the top and East is to the left. \label{fig:ident} Uncertainty
in the placement of the slit is estimated to be within $\pm$0.5"}
\end{figure}

\begin{figure}
\plotone{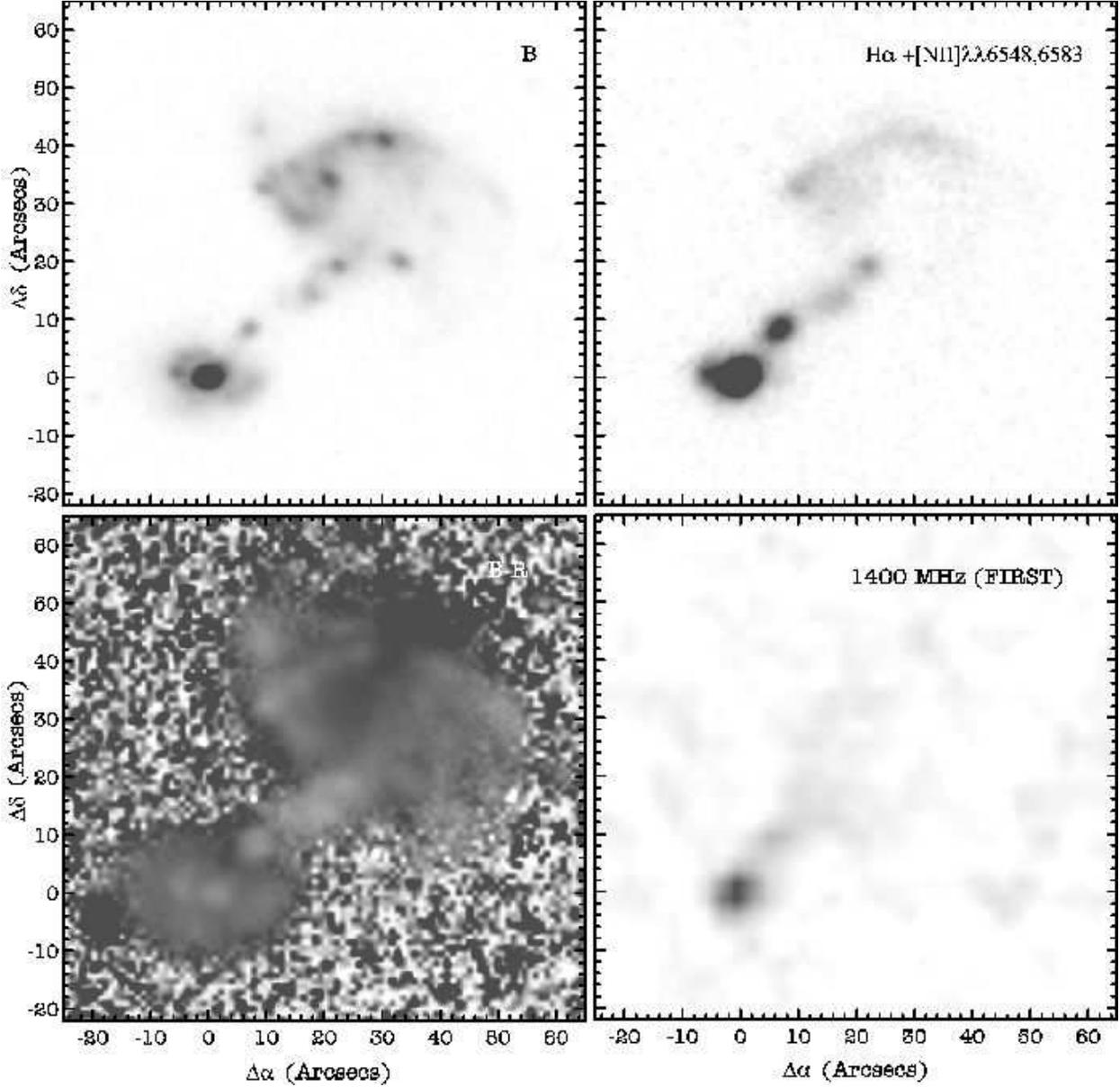} \figcaption[plotall.eps]{Multifrequency
morphology of the Arp 194 system. Upper left panel: Johnson B image. Upper
right panel: Continuum-subtracted narrow band image [\ha+\nii]. Lower left
panel: {\sl B -- R} color map. Limits on scale are -1.2 (white) and 1.2
(black). Lower right panel: radio emission at 1.4 GHz from FIRST data. In
all panels, North is to the top and East is to the left.
 \label{fig:morph} }
\end{figure}

Arp 194 (Arp 1966;  $\equiv$ UGC 06945 $\equiv$ VV 126;
Vorontsov-Vel'Yaminov 1977) is a small-angular-size system of two major
components:  a  north-western  galaxy of disrupted morphology (0'.8 x
0'.6;  see Fig. \ref{fig:ident} and Fig. \ref{fig:morph}), and an apparently
more regular galaxy (0'.35 $\times$ 0'.35) located $\approx$ 40 arcsec to
the South-East.  From the heliocentric velocity of the Arp 194S ($\equiv$
A194S hereafter), \vr $\approx$ 10500 \kms\ (see \S \ref{res}), we infer a
distance of 175 Mpc for the system (H$_0 \approx$  60 \kms\ Mpc$^{-1}$).
Arp 194N  and Arp 194S are therefore separated by a projected linear
distance $d_{\rm p} \approx$ 34 Kpc (1 arcsec corresponds to $d_{\rm
p}\approx$ 850 pc). Arp described the Arp 194 system as belonging to the
class of ``galaxies with material ejected from nuclei.'' Arp noted further
``outer material connected by thin filament to very hard nucleus.'' Several
peculiar features of Arp 194N ($\equiv$ A194N) have been also discussed in
an early study (Metlov 1980).

Theoretical modeling involving a reliable treatment of dissipative
phenomena like star formation, as well as  high resolution numerical
simulations of stellar and gas motions are still the current frontier in
our understanding of interacting galaxies (see e.g., Hearn \& Lamb 2001;
Semelin \& Combes 2000; Barnes \& Hernquist 1998; Barnes \& Hernquist 1996;
Mihos \& Hernquist 1996).  From simulations, we expect a wide range of
phenomenological properties due to  interaction. Gravitational forces can
marginally enhance the star formation rate (SFR) in an unbound
intergalactic encounter as well as give rise to the most luminous bursts of
star formation observed in merging systems (see e.g., Krongold et al. 2002,
and references therein).  Tidal effects drive noncircular motion in disk
galaxies, can strip a significant amount of stars and gas, and even lead to
the formation of the so-called tidal-dwarf galaxies (Barnes \& Hernquist
1992). A related issue is the role of gravitational interaction in fueling
nonthermal nuclear activity. Both simulations and observations of single
interacting systems are a necessary complement to statistical studies on
the frequency of interacting systems among active galaxies, since the latter
provide valuable but only circumstantial evidence in favor of interaction
as a major driver of nuclear activity (Krongold et al. 2002; see however
Schmitt 2001, and references therein).  Collisional ring galaxies are
excellent laboratories for studying galactic evolution, global star
formation, and the occurrence of gas cross-fueling among galaxies.  The
prototype of this class of objects is the Cartwheel galaxy A0035-33. Only
few systems similar to this object are known: among the best studied we
recall VII~Zw~466, Arp 10, II Zw 28, II Hz 4, Abell 76, LT 36, NGC 985, LT
41 (Appleton \& Marston 1997).  Other likely cases  are Arp 119 (Hearn \&
Lamb, 2001), Arp 118 (Charmandaris et al. 2001), Arp 284 (Smith \& Wallin
1992, Smith et al. 1997), and Arp 143 (Appleton et al. 1992). The dominant
ring morphology is thought to result from a head-on collision between two
galaxies, one of which  traveled close to the spin axis of the other,
striking the disk close to its center.  The resulting gravitational
perturbation is believed to drive a set of symmetrical waves or caustics
through the stellar disk (Lynds \& Toomre 1976, Theys \& Spiegel 1976,
Toomre 1978, Appleton \& Struck-Marcell 1987, Struck-Marcell 1990).
Computer simulations of such head-on encounters, in which at least one
galaxy has a significant gaseous component, show indeed the production of
density enhancements and shock waves in the interstellar medium. These high
density regions coincide with the location of recent, large scale star
formation observed in  ring galaxies.

Fig.  \ref{fig:ident} and Fig. \ref{fig:morph} suggest that A194N is a
collisionally  induced ring galaxy connected to the past intruder A194S by
a string of relatively bright knots (\S  \ref{res}).  Several photometric
properties indicate star formation rates typical of Starburst galaxies (\S
\ref{sform}). The surprisingly strong radio emission from A194S and knot A
is likely due mainly to supernova remnants which imply even higher star
formation rates (\S \ref{rsn}) and substantial internal obscuration.
Relatively rare systems like Arp 194 where gas motions are induced by a
head-on encounter enable us to unambiguously infer the geometry of the
system. In this case that information provides evidence for
``cross-fueling'' (see \S \ref{int}).  Our results for Arp 194 have
intriguing implications  on the analysis and interpretation of galactic
superwinds (\S \ref{disc}).

\section{Observations, Data Reduction \& Analysis \label{data}}

Johnson (B, R) and narrow band (\ha) images of Arp 194 were obtained at the
Cassegrain focus of the 2.1 m telescope of the Observatorio Astr\'onomico
Nacional of M\'exico at  San Pedro Martir (SPM) on January 30--31,  1995.
 The detector was a Tektronik CCD (24 $\mu$m square pixel
size, 1024$\times$1024 format) that yielded a scale of 0.315" pixel$^{-1}$,
and a field of view of 5.4'.  A narrow band filter with $\lambda_C \approx
6819$ \AA, and $\Delta \lambda \approx 86$ \AA\ allowed coverage of the
\ha+\nii\ blend of Arp 194 (\S \ref{multi}). Table \ref{tab:obs} provides a
log of each single exposure taken at SPM.

Long slit spectra were obtained on February 2, 1995 at the 2.1 m telescope
equipped with a Boller \& Chivens spectrograph. A 1200 lines mm$^{-1}$\
grating was employed for observations of the \ha\ spectral region (coverage
$\approx$ 6170 -- 7240 \AA). Four spectrograms were taken, two at position
angle P. A. $\approx$ 145$^\circ$ and two P. A. $\approx$ 118$^\circ$, with
an exposure time totaling 60 min. at each P. A. (see Fig. \ref{fig:ident}
for slit placement). The slit  was opened to 200 $\mu$m (2.6"  on the focal
plane of the telescope). This setup resulted in a resolution of $\approx
2.5$ \AA\ FWHM ($\approx$ 110 \kms\ at redshifted \ha).  The data reduction
for both imaging and spectroscopic data and the calibration procedure
followed standard IRAF practice and has been identical to that employed by
Marziani et al. (1999), who used data obtained during the same nights. We
will not describe them again here. Table \ref{tab:spec} presents line
fluxes for emission line regions isolated along the slit.


\section{Results \label{res}}

\subsection{ Morphology \& Photometry \label{multi}}

The morphology of the Arp 194 system in the B band is shown in Fig.
\ref{fig:ident}, where some of the most prominent features are identified.
A multifrequency view is provided in Fig. \ref{fig:morph}. The four panels
show (clockwise from top left) the SPM B-band image (with a cut optimized to
show the internal structure at the expense of loss of the fainter
envelope), the continuum subtracted \ha\ + \nii\ narrow band image, the
B--R color index map, and the radio map at 1.4 GHz obtained from the FIRST
survey (Becker et al. 1995).  Table 2 reports the photometric properties of
the main features visible in these maps. The uncertainty in the B and R
photometry is estimated to be $\approx \pm 0.05$ mag at a 2$\sigma$\
confidence level. B--R values are therefore accurate within $\la$ 0.1 mag.
The uncertainty associated to the overall \ha + \nii\ calibration is
estimated to be within 10 \%. Due to the diffuse nature of large part of
\ha\ emission, an appropriate uncertainty is probably a factor 2 for the
measurements on the faintest and most diffuse regions, while it should be
within 20 \% for the brightest and least diffuse ones. A194S is  a
surprisingly strong radio source when compared to nearby Seyfert galaxies.
The specific power of A194S is $\log P_{\nu,1.4 GHz} \approx$ 22.1, with
$P_{\nu, \rm 1.4 G Hz}$ in W Hz$^{-1}$), placing it at the high end of the
distribution of radio power in Palomar Seyferts, which cover the range
$\log P_{\nu, \rm 1.4 GHz} \approx$ 18--22 (Ulvestad \& Ho 2001; Ho \&
Ulvestad 2001). Of course this is still negligible compared to nearby radio
galaxies such as NGC 1275 with $\log \nu P_{\rm \nu,1.4 GHz} \approx$ 25.6
(Owen et al. 1980).

Another remarkable feature of this system is a string of relatively bright
knots (``blobs'') in between the two components of Arp 194, aligned along
the north western direction (the ``streamers'' in the terminology of Metlov
1980). The blobs B and C have B--R$\approx$ 0.12 and 0.26 respectively.  In
Table \ref{tab:mf}, the J and K magnitudes are upper limits from the
isophotal contours published by Bushouse \& Stanford (1992). The blobs are
not present at the lowest contour levels traced by these authors. The trail
of blobs between A194S and A194N can be readily interpreted as stripped gas
due to the interpenetrating encounter between A194S and A194N (see Marziani
et al. 1994; Horellou \& Combes 2000). There are several lines of evidence
in favor of this interpretation:
\begin{itemize}
\item the rough alignment of the three main blobs observed in between A194S and
A194N. The blobs literally provide a trail of   ``footprints''  tracing the
path of the intruding galaxy (A194S) after crossing  A194N. They identify a
direction pointing close to the geometrical center of A194N;
\item the lack of any K counterpart for all blobs, which  suggests a common
origin for the blobs. In particular, it rules out that blob A is a
background galaxy;
\item almost no solution of continuity nor any radial velocity discontinuity  \ha\
emission in the 2D spectra at P. A. 145$^\circ$\ (see Fig. \ref{fig:pa145bw}
and \S \ref{rv});
\item several features of A194N, which can can be ascribed to a collisional ring, as discussed below.
\end{itemize}

Following the collision interpretation we might expect Arp194 to show
features predicted by models for off-center interpenetrating encounters
including: (1) a prominent outer ring and (2) displacement of mass toward
the ring (Gerber et al. 1992; see also the K images of several ring
galaxies in Appleton \& Marston 1997). A194N can be interpreted as a
distorted ring galaxy although tracing the outer ring is not easy. Two arcs
towards the south and north can be traced unambiguously, suggesting a ring
of radius $R_{\rm d} \approx$ 17 Kpc. The hollow region towards the SE
reinforces the collisional ring galaxy interpretation because it cannot be
easily explained in other terms. Another supporting element involves the
trend in radial velocities along P.A.=145$^\circ$, which can be more easily
explained as the result of expansion than rotation (see \S \ref{rv}). A194S
shows a larger and brighter nuclear bulge suggesting that it is an earlier
morphological type (Sa-Sb?) than N (originally type Sc?). Two blue arc-like
features (Fig. \ref{fig:ident} \& Fig. \ref{fig:morph}) are displaced
towards the eastern and western sides of the A194S nucleus (see also the
color map). They may be signatures of a ring-like expansion wave produced
by the assumed  head on encounter between A194N and S.

Interpretation of A194N is complicated by the complex morphology on the
north-western side. Two main condensations (corresponding roughly to A194N-A
and A194N-B) are visible in the K images. This may indicate that we are
detecting the nucleus of N and another galaxy projected nearby (i.e. A194N
+ an additional perturber). The color map reddest region to the North
($B-V\ga 1$) may be associated to such a perturber.  The issue requires
more data but the trail of blobs suggests that A194N and S are the main
players in this interaction. Our discussion reflects that assumption.

\subsection{Emitting Regions \label{ler}}

\begin{figure}
\plotone{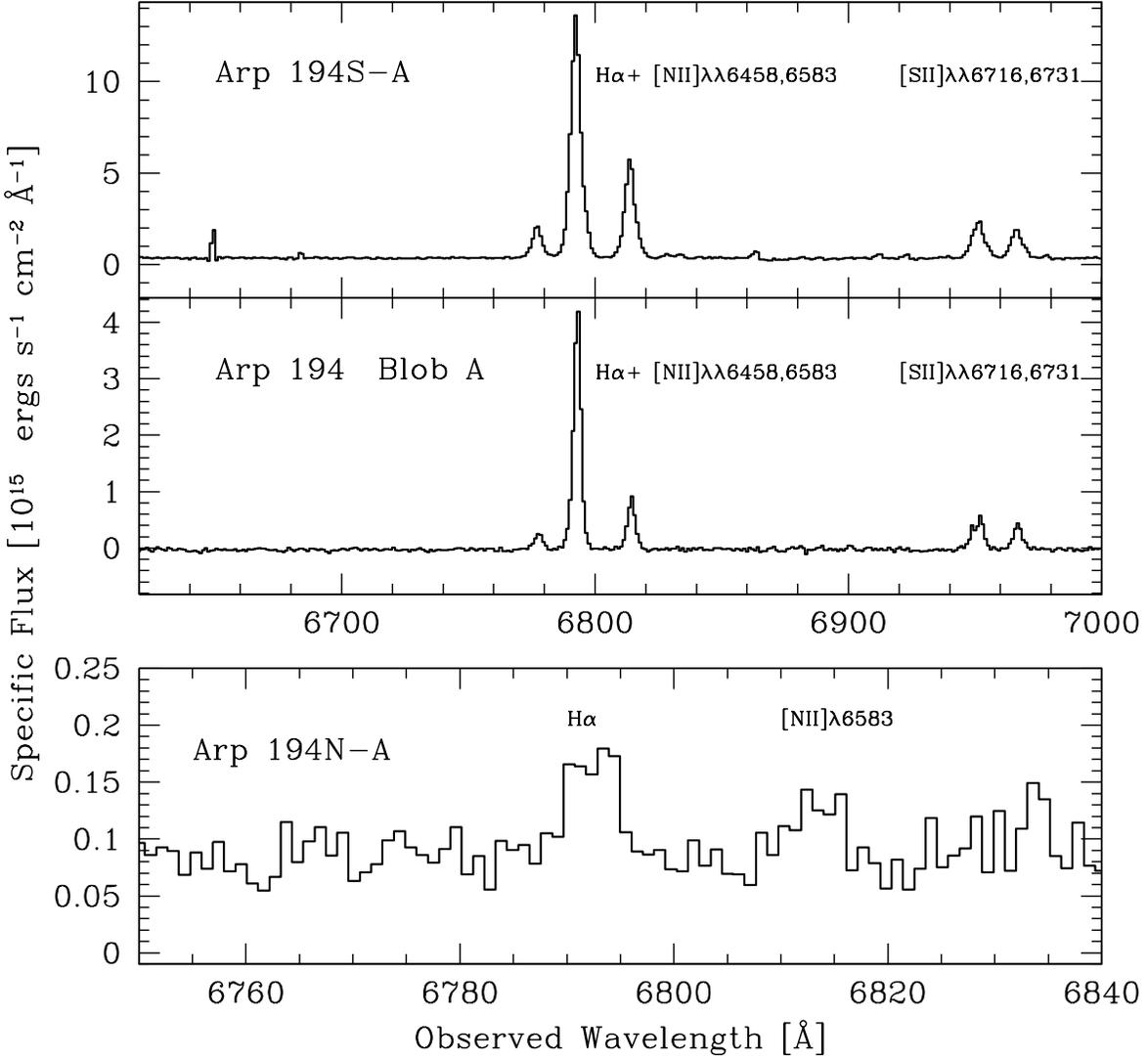} \figcaption[spectran.eps]{\ha\ spectral
regions of relevant emitting line regions, ordered from the South to the
North. Abscissa is observed wavelength in \AA, ordinate is specific flux in
units of \ergss cm$^{-2}$ \AA$^{-1}$. The A194N-A spectrum (bottom panel)
spectral range has been expanded, to better show the boxy and broad
appearance of \ha. \label{fig:spectran}}
\end{figure}

\paragraph{The Nucleus of A194S $\equiv$A194S-A} The intruder nucleus
is a luminous \ha\ source (L(\ha) $\approx$ 2$\times 10^{41}$ \ergss) with
color index B--R$\approx$0.5. The ratio I([\ion{N}{2}]\l 6584)/I(\ha)
$\approx$ 0.4 (Fig. \ref{fig:spectran} and  \ref{fig:crossdisp}) suggests
that there is no significant source of ionization other than hot stars. The
moderate I(\sii)/I(\ha) ratio ($\approx$ 0.3) and very low I([\ion{O}{1}]\l
6300)/I(\ha) ratio ($\approx$0.02) support this conclusion. A further
confirmation comes from the I([\ion{O}{3}])\l 5007/I(H\b) ratio ($\approx$
0.8) measured by Metlov (1980).  Line ratio diagnostic diagrams (Veilleux
\& Osterbrock 1987) show that A194S is located very close to the \hii\
zone. Our data show no evidence for non-thermal nuclear activity. In this
respect, it is worth noting that there is (Fig \ref{fig:crossdisp}) no
variation of the I(\ha)/I(\nii) ratio along the slit (an increase would be
expected if a nonthermal contribution were present). The ratio
I(H$\alpha$)/I(H$\beta$) measured by Metlov (1980) 3.7:1.0 suggests
moderate extinction ($E(B-V)\approx0.2$). The heliocentric \vr\ $\approx$
10502 \kms\ agrees with Metlov's (1980) determination within 10 \kms.

\paragraph{Blob A} This is the second strongest source of \ha\ emission
in the interacting pair and it is also clearly detected as a FIRST radio
continuum source.  Metlov (1980) measured I(\ion{O}{3})\l 5007/I(H\b)
$\approx$ 1.56. The intensity ratios I(\ion{N}{2}\l 6583)/I(\ha) $\approx$
0.23, I(\sii)/I(\ha) $\approx$ 0.24, and I(\ion{O}{1})\l 6300/I(H\a)
$\approx$ 0.027 indicate that \hii\ emission is dominant and this is
confirmed by the source location in the diagrams of Veilleux \& Osterbrock
(1987). A ratio L(\ha)/$\nu$P$_{1.4 GHz} \approx$ 800 is a useful measure
of what can be inferred about star formation processes. It is noteworthy
that Blob A is clearly connected to the nucleus of A194S by more extended
line emission (Fig.  \ref{fig:pa145bw}; see \S \ref{rv}).

\paragraph{A194N-A}  Source A194N-A is  likely to be the nucleus of the
disrupted spiral but it was only partially in the slit at both position
angles. \ha\ and \nii\ show FWHM $\approx$ 250 \kms which is approximately
twice as broad  as the adjacent emitting regions.  The observed broadening
is significant at a confidence level $\simgt$ 3 $\sigma$ taking into
account a FWHM error of 15\% or $\approx 30$ \kms (1$\sigma$). The
broadening is appreciable in both spectra and it is  due to the presence of
an additional (redshifted) line component in the slit since  the \ha\ and
\nii\ profiles  are double-peaked with peak \dvr\ $\approx$ 130 \kms\ (see
Fig. \ref{fig:spectran}, where spectra of the main emitting regions are
shown). The I(\nii)/I(\ha) ratio ($\approx$ 0.6) for the red component
suggests the presence of non-thermal emission probably associated with
shock-heated gas (not strong, otherwise the \sii\ lines would be detected).

\subsection{Radial Velocity Curves \label{rv}}

\begin{figure}
\plotone{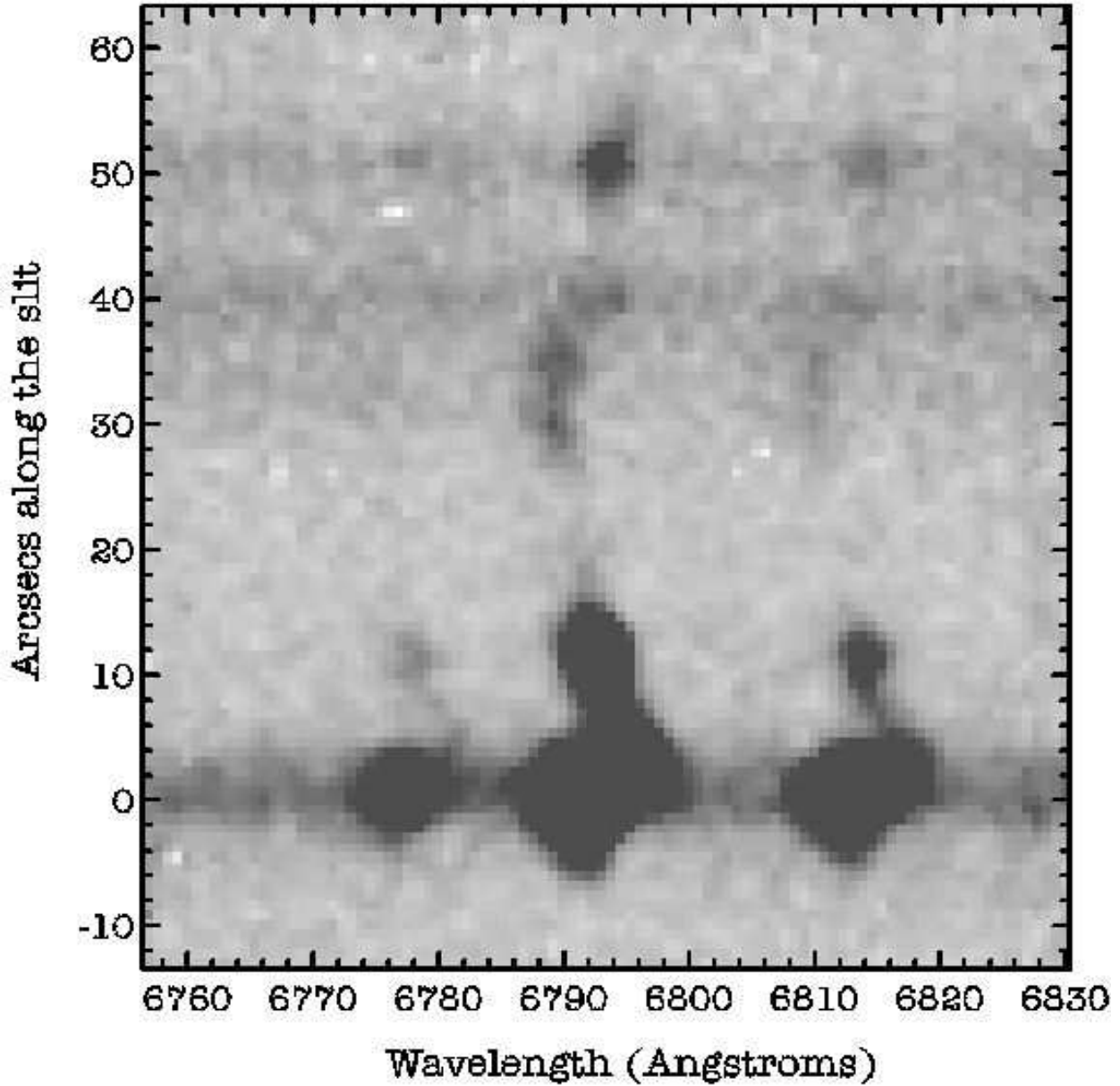} \figcaption[arp194pa145.eps]{Long slit spectrum
of Arp 194, at P. A. $\approx$145$^\circ$, in the range covering \ha\ and
\nii. Abscissa is observed wavelength in \AA ngstroms; ordinate is arcsec
along the slit (NW to the top). The light contour levels are meant to
emphasize the line structure in the inner few arcsecs of A194S.
 \label{fig:pa145bw}}
\end{figure}

\begin{figure}
\plotone{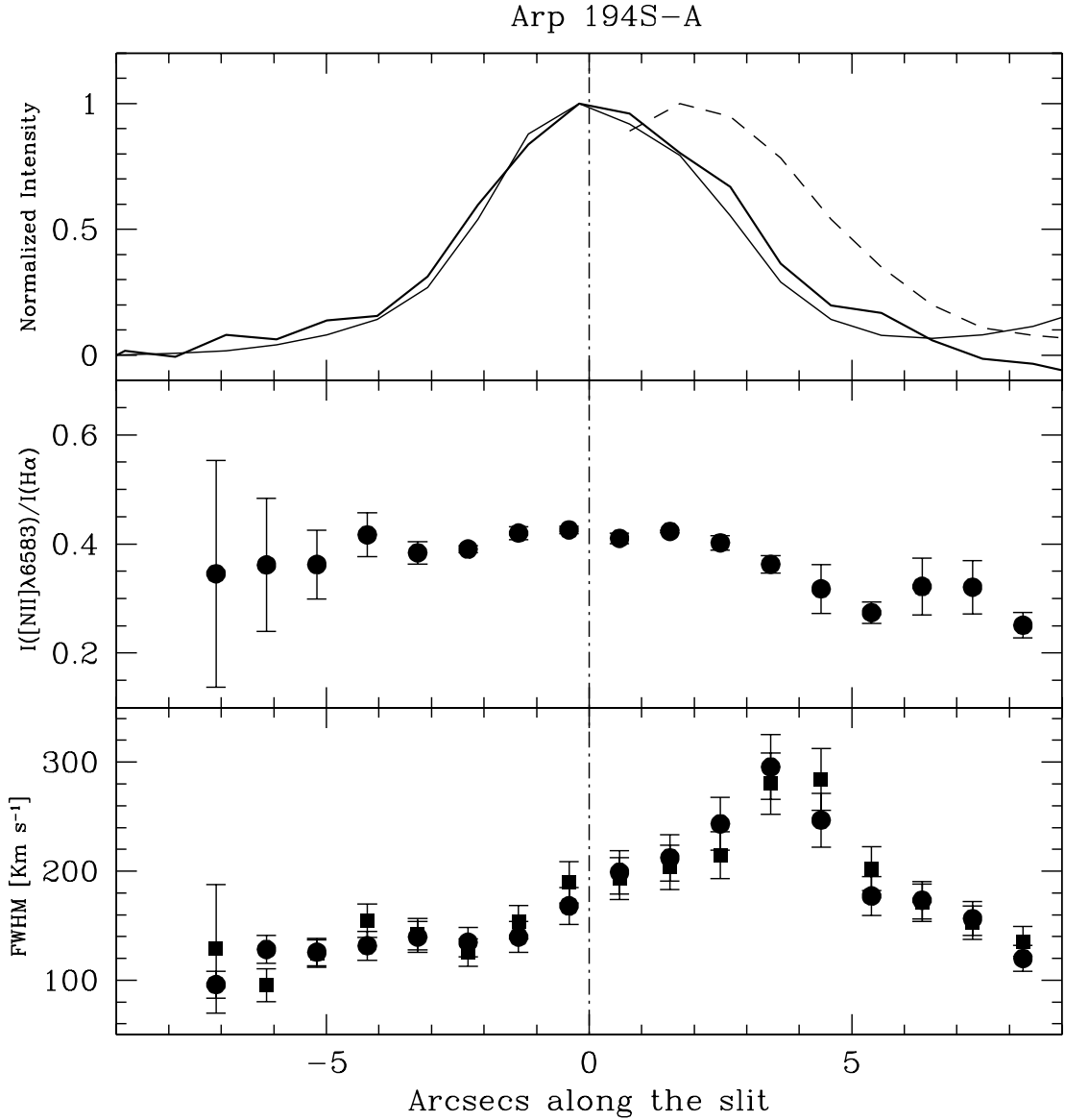} \figcaption[crossdisp.eps]{Upper panel: cross
dispersion intensity profile for \ha\ emission (thin solid line), continuum
(thick solid line; assumed to peak at the abscissa 0 point), and extended
\ha\ component (dashed line). Ordinate is intensity normalized to peak
value. The extended \ha\ component peaks approximately 2" from the
continuum, and it is barely visible in the strip of maximum continuum
emission (see also \ref{fig:pa145bw}). Middle panel: Intensity ratio of the
\ha\ and [\ion{N}{2}]\l 6583 lines, as  a function of the distance from the
nucleus of A194S. Lower Panel: for FWHM(\hb) (filled circles) and
FWHM([\ion{N}{2}]\l 6583) (filled squares) vs.  distance from the nucleus
of A194S. Ordinate is FWHM in \kms\ corrected for instrumental profile.
\label{fig:crossdisp}}
\end{figure}

\begin{figure}
\plotone{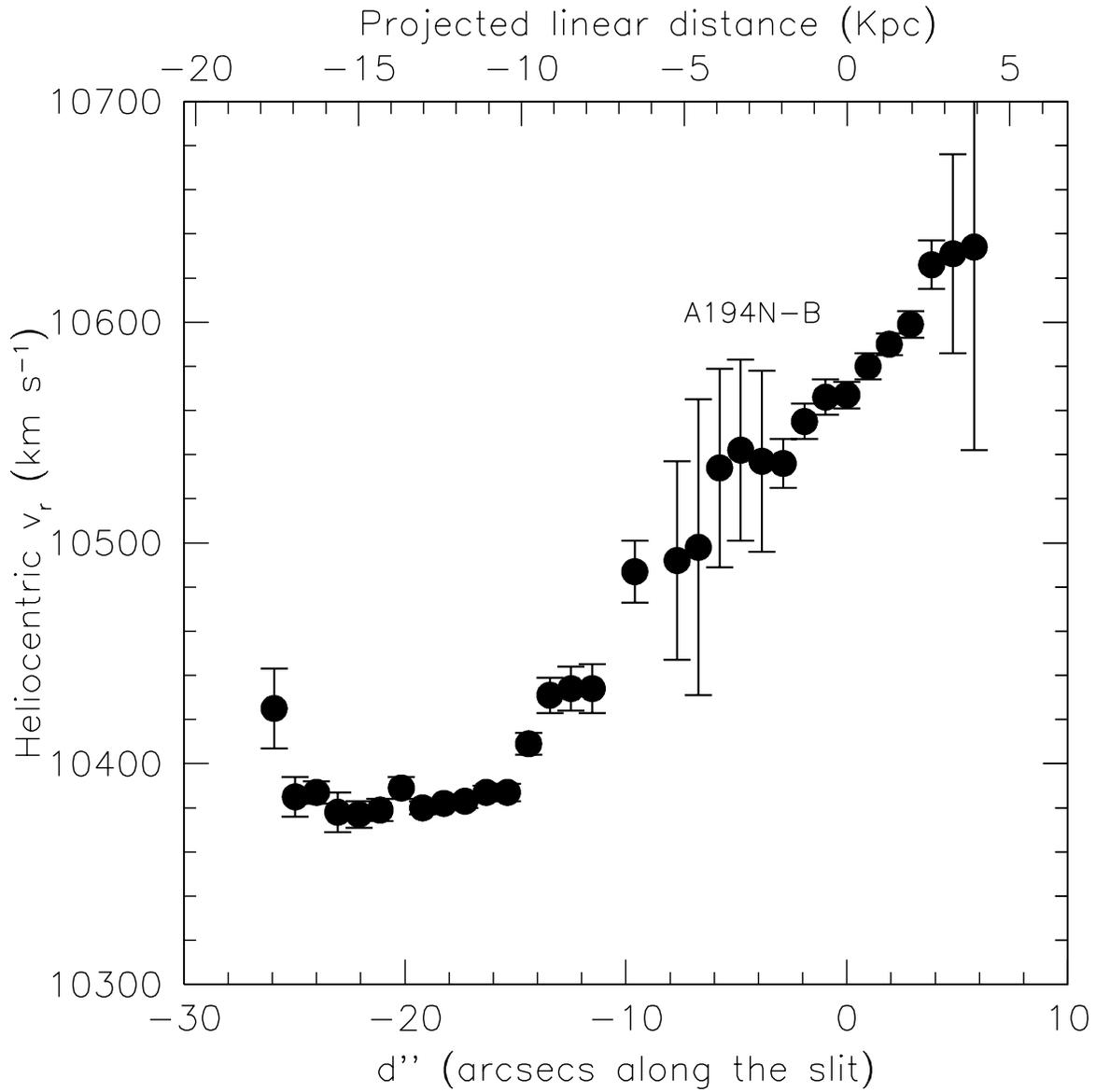} \figcaption[arp194pa118.eps]{Heliocentric
radial velocity curve of the Arp 194 system at P. A. = $118^\circ$. Note
that, at variance with the previous figure, the origin has been set on
A194N-B. \label{fig:pa118}}
\end{figure}

Fig. \ref{fig:pa145bw} is a grey-scale reproduction of the  extended \ha\ +
\nii\ emission. The   radial   velocity   (\vr)   curve  derived from this
figure is shown in Fig. \ref{fig:pa145} (centered on A194S-A) and in Fig.
\ref{fig:pa118} at  P.  A.  $\approx$ 118$^\circ$. Both radial velocity
curves agree qualitatively with the data provided by Metlov (1980).
Emission is continuous between blob A and A194S.  \vr\  increases
monotonically to within  $\approx$ 3"  NW of A194S-A (maximum $\Delta
v_{\rm r}$ $\approx$ 120 \kms). This emission line component produces the
cusp in the radial velocity curve between 0--5" from  the continuum peak of
A194S. The velocity curve on the northern side of A194S is obviously not
due to rotational motion because it appears to turn over.  The velocity
cusp is most likely due to blending of the rotational velocity field with
the redward-displaced component visible in Fig. \ref{fig:pa145bw}. This is
made more evident by the contour overlaid on Fig. \ref{fig:pa145bw} and by
the cross-dispersion intensity profile shown in the upper panel of Fig.
\ref{fig:crossdisp}. The main \ha\ component has been summed over
$-160$\kms $\la \Delta v_{\rm r} \la$110 \kms; the redward component in the
range $110$ \kms\ $\la \Delta v_{\rm r} \la$ 330 \kms. In correspondence of
$\Delta d" \approx 4"$, the total \ha\ emission is significantly broader
since the two \ha\ components are blending together (the \ha\ profile
appears double-peaked).

\subsection{The Geometry and Kinematics of the Encounter}

\begin{figure}
\plotone{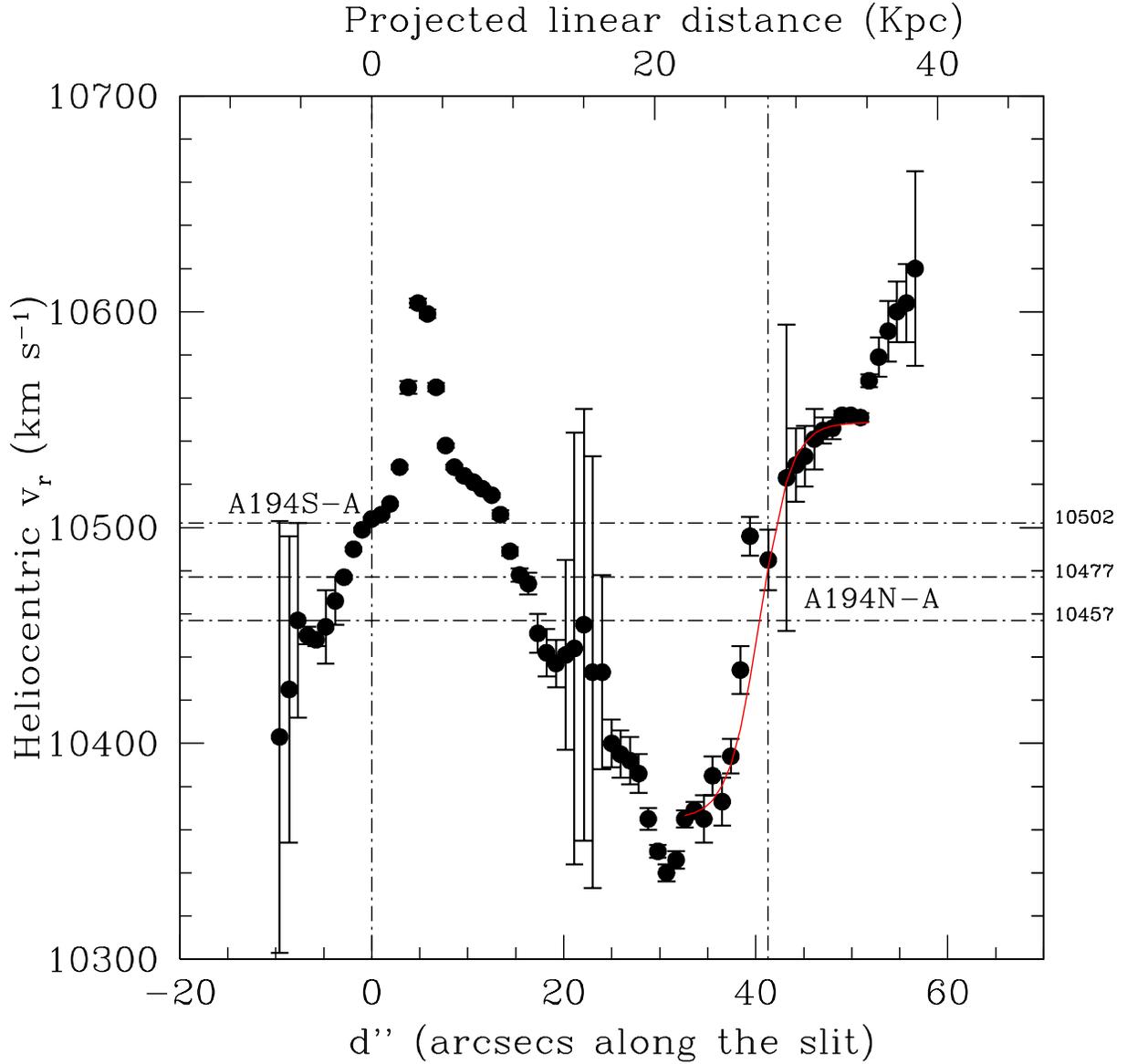} \figcaption[arp194pa145.eps]{Heliocentric
radial velocity curve of the Arp 194 system at P. A. = $145^\circ$. Note
that the apparent discordant point at $\approx$ 40'' reflects a broadening
of the \ha\ and \nii\ lines. The origin of the projected linear and angular
distance scale has been set coincident with A194S-A (at cross-dispersion
peak of the continuum). The three dot-dashed lines draw the systemic \vr\
measured for A194S (\vr $\approx$10502 \kms) and A194N (\vr\ $\approx$
10457 \kms\ and 10477 \kms; see text for details).
 \label{fig:pa145}}
\end{figure}

Any  inferences on the systemic \vr\ of A194N-A are uncertain (\S
\ref{ler}). The P.A. = 145$^\circ$\ slit, however, crosses A194N not far
from the geometrical center of the ring.  Our spectrum at PA= 145$^\circ$
mainly samples the ``empty'' inner region of A194N. This gives us confidence
that we are measuring a reasonably reliable radial velocity $v_{\rm r,N}$\
for A194N at $\approx$ 41" in Fig. \ref{fig:pa145} (the edges of the ring
are at 30" and 50", and interestingly, in their correspondence the \vr\ is
constant over 3"). A fit to the \vr\ curve in this region is shown by a
filled thin line in Fig. \ref{fig:pa145}. We derive a  value $v_{\rm r,N}
\approx$ 10477 \kms if we consider the midpoint between the two segments of
constant \vr; $v_{\rm r,N} \approx$ 10457 \kms\ if we consider the average
\vr\ between the two segment. The roughly symmetric appearance of the radial
velocity curve around this point reinforces our confidence that $v_{\rm
r,N}\la$ 10500 \kms. We derive from the observed continuum peak (the 0
point of Fig. \ref{fig:pa145}) $v_{\rm r,S} \ga$ 10502 \kms. Therefore,
A194S is receding from A194N with  $\Delta v_{\rm r, coll} \approx 25 - 40$
\kms.

We conclude that A194S is more distant  and this implies that the motion of
the blobs relative to it follows straightforwardly. Blob A, in the 2D
spectrum of Fig. \ref{fig:pa145bw} is connected to the emission line
component of increasing \vr\ detected up to a few arcsecs from the
continuum peak. The \dvr\ between A194S-A (\vr\ measured at continuum peak)
and blob is still positive ($\Delta \rm v_r \approx 30 $\kms). Therefore,
blob A cannot be the product of outflow for the obvious reason that it
would be moving in the wrong direction. {\em  If the blobs trail A194S,
Blob A -- as well as the extended emitting gas of increasing $v_r$ in
between blob A and A194S-A -- must be falling toward A194S-A.} The validity
of this results is based on two considerations: (1) the blobs, and blobs A
especially, trail after A194S; (2) $\Delta v_{\rm r,coll} \ga 0$ \kms,
implying that A194S is further from us than A194N. In this case, blob A is
closer to us than A194S. We must remark that, albeit $\Delta v_{\rm
r,coll}$ is small, $\Delta v_{\rm r,coll} \la 0$ \kms\ is inconsistent with
the \vr\ curve and the geometry of the system (see Fig. \ref{fig:pa145}).

\paragraph{Ring Age and Time After Crossing} To  estimate  the  time after crossing,
we can  consider that, in the special case of a head-on encounter in which
the direction of motion of the intruder is strictly perpendicular to the
plane of the target galaxy, the distance between the intruder  and the
target  galaxy can be deprojected once  the inclination  of the  target
galaxy is known (this may not be strictly true for Arp 194; this
 is just a first approximation). The main underlying
assumption is that the ring starts to expand at the time the intruder
nucleus was interpenetrating the disk of the target, as suggested  by
theory  and numerical simulations (Lynds and Toomre 1977). The expansion
velocity will be $ v_{exp} = R_{\rm d}/d_{\rm p} \Delta v_{\rm r} \tan i
\approx 25 \Delta v_{\rm r,coll,50} \tan i$ \kms. The time after crossing
is $\tau = R_{\rm d} / v_{\rm exp}$. We can estimate the inclination of the
ring making two extreme choices: (1) the NW end is traced by the Arc Spots
A,B,C; (2) the NW end is the fainter spot $\approx$5" to the NW from the
Arc. Since this implies $54^\circ \la i \la 75^\circ$, we obtain
2$\times$10$^8$\ yr$ \la \tau \la 6\times10^8$ yr if $\Delta v_{\rm r,coll}
= 40$ \kms. If $i = 64.5^\circ$, $\tau \approx 4\times 10^8$ yr.

\subsection{Star Formation Properties \label{sform}}

The total \ha\  luminosity of the Arp 194 system is $\approx 1.5 \times
10^{42}$ \ergss.  This yields a star formation rate $\approx$ 10 \msol
yr$^{-1}$ for masses between 0.1 and 100 \msol, assuming a Salpeter Initial
Mass Function (see Kennicutt 1998 for relationships  and references).  This
value   is   comparable to the one  of   powerful Starburst
galaxies.\footnote{Unfortunately,  Arp 194 has not been covered by IRAS
observations.} It is interesting to note that if we consider the tight
FIR-radio correlation of star-forming systems (Mirabel \& Sanders 1996),
with logarithmic index q=2.35, $\log$~ $L_{\rm FIR} \approx $ 14.92 $\log
P_\nu$, we obtain $L_{\rm FIR} \approx 2.3 \times 10^{44}$ \ergss\ $\approx
6\times 10^{10} L_\odot$. Arp 194 may be well a luminous IRAS galaxy. A
vigorous star formation going on in the produced ring seems to be a common
property of ring galaxies (Marston \& Appleton 1995).  Optical images often
show a compact off-centered nucleus surrounded by a number of blue knots
that are HII regions. The ring component often hosts giant molecular
complexes, and its color is consistent with a very young stellar population.
Indeed, the ring arcs of A194N stand out in the color map since they are
bluer than the surrounding regions.

\paragraph{Is it just a Starburst?}
A  Starburst surrounding the nuclear region, as well as infall of gas from
above the galactic plane may produce  extinction to the point to fully
obscure an active galactic nucleus (AGN).   Actually, the early stages in
the life of an AGN may be dominated  by obscuration,  so that  even the
very AGN detection may be troublesome.  This seems to be true over a wide
range of luminosity, from ultra-luminous far  IR galaxies ($L_{\rm FIR} \sim
10^{12} L_\odot$), down to the nearest Seyfert 2 (see e.g., Krongold et al.
2002; Maiolino et al. 2000; Dultzin-Hacyan 1995; Sanders et al. 1988). We
computed the $L$(\ha)/$\nu P_\nu$\ at $\nu = 1.4$ GHz as a function of the
aperture radius for increasing apertures from 2" to 9". In this interval
the ratio is $L$(\ha)/$\nu P_{\nu} \approx$ 2200, with a $\pm$ 10 \%\ change
along different aperture sizes. The nuclear value is larger than the value
observed on Blob A (see below). Blob A is not strictly a galaxy and most
likely  lacks the deep potential well associated to the nuclei of galaxies
where massive black holes are thought to be present. Furthermore, both the
nucleus and blob A obey to the correlation between $f$(\ha+\nii) and radio
$f_\nu$\ at 20 cm found by Kennicutt (1983) for spiral galaxy, which
implies  $f$(\ha+\nii)/$ \nu f_\nu \approx 1500$. This suggests no
contribution from an obscured active galactic nucleus in A194S-A.

Having  ascertained that most of the \ha\ and radio emission is due to \hii\
regions, we can deduce several parameters related to star formation
reported in Table \ref{tab:sfr}. We will focus on blob A and B for which we
have a more complete dataset. The absence of any J, and K counterparts for
the blobs (Bushouse \& Stanford 1992) readily suggests that, if the blob are
associated to star forming regions,  their age must be relatively young,
and that mainly gaseous matter has been stripped through the encounter.

\begin{figure}
\plotone{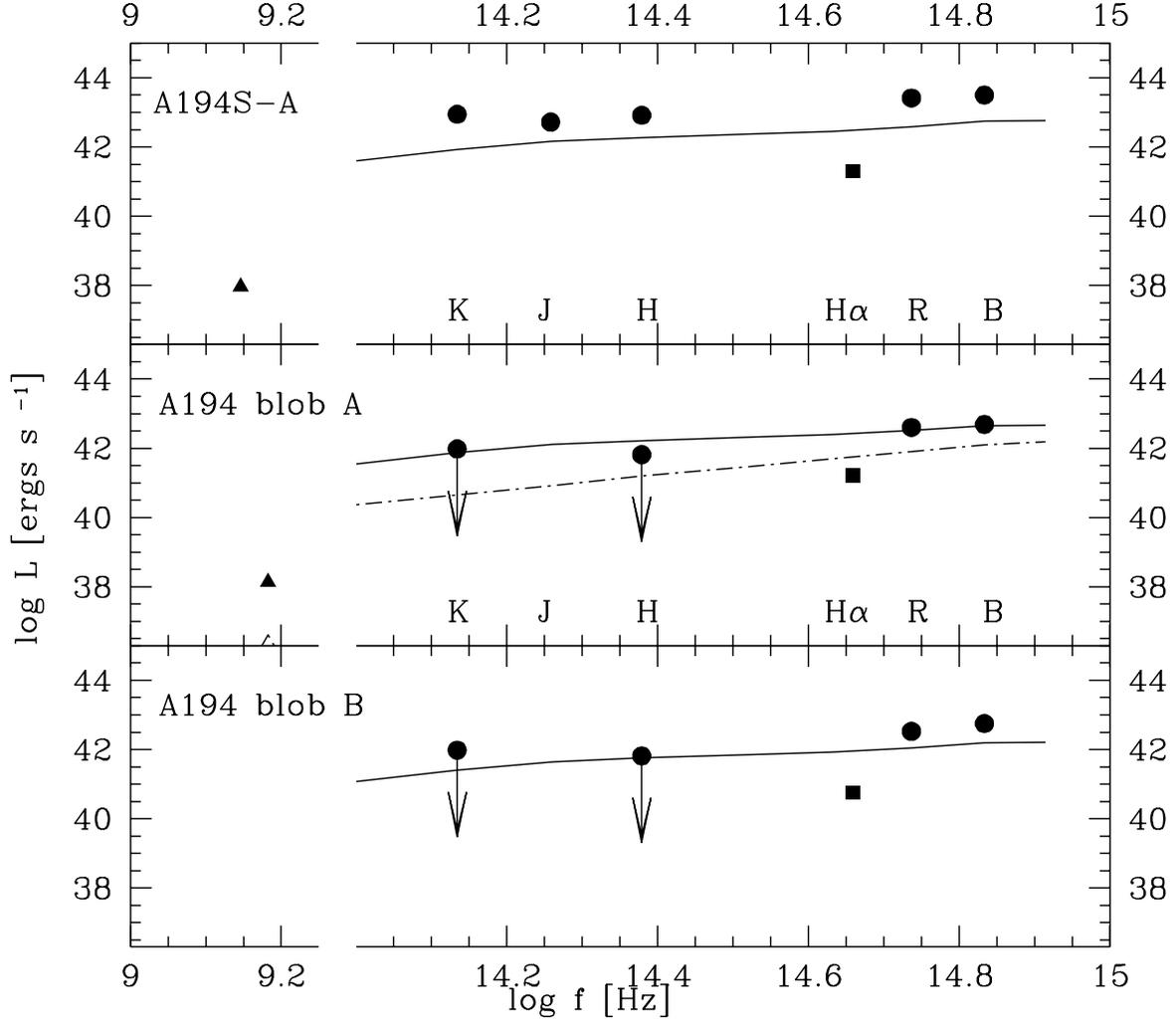}
 \figcaption[fig09_a194.eps]{Photometric properties of A194S-A, blob A and
blob B and population synthesis. Abscissa is logarithm of frequency in
Hertz, ordinate is logarithm of $\nu L_\nu$\ in ergs s$^{-1}$. The filled
squares represent the \ha\ luminosity. The solid line shows the prediction
of {\tt STARBURST99} simulations for the continuous SFR as reported in the
4th column of Table \ref{tab:mf}: 1.55, 1.2, 0.42 \msol yr$^{-1}$ for
A194S-A, blob A and blob B respectively, after a time $\approx$7$\times
10^7$ yr. Errors are approximately the size of the symbols or smaller. The
dot-dashed line of the middle panel is a simulation for the ``first
generation of stars'' hypothesis (see text) with a main sequence total mass
${\cal{M}}_{\rm tot} \approx 5 \times 10^6$ \msol. The triangle refer to
the expectation for radio power in case of only free-free emission. No
correction for external extinction has been applied. \label{fig:sb99}}
\end{figure}

\paragraph{``First  Generation''  of Stars in  Blob A?} If we were  observing
 a very young  population (i.e.,  a first  generation  of  stars in which
 {\em all} stars are still in the main sequence with a total mass reported in Col. 7 of Table
\ref{tab:sfr}), radio emission should be due exclusively to thermal
Bremsstrahlung in the \hii\ regions. In this case, the emissivity ratio
between \ha\ and radio specific flux is dependent only on the electron
temperature $ T_{\rm e} $\ (Osterbrock  1989,  pp. 80, 88, 95). Assuming $
T_{\rm e} \approx$ 10$^{4\circ}$K,  we  obtain that the expected ratio  at
$\nu=$ 1.4 GHz is $j_{\rm H\alpha}/\nu j_\nu \approx $44000 for optically
thin free-free emission.  This value is most likely an underestimate
because the emitting region  should be optically thick at 1.4 GHz: \hii\
regions have frequently a turnover  frequency between the optically thick
and thin regime at $\approx$ 3 GHz. Since the observed ratio is more than an
order of magnitude smaller, we conclude that radio emission must be mainly
due to a non-thermal sources associated to supernov\ae. The ``first
generation" scenario is not supported  by the  optical properties either.
We computed the numbers of ionizing photons needed to sustain the L(\ha),
and hence the number of OB stars (spectral type earlier than B9) under the
assumption that no photon escapes from the nebulae. Approximately  30000 OB
stars are needed for blob A. However, a larger number of OB stars would be
needed to account for the B absolute magnitude ($M_{\rm B,blob  A}\approx
-$17.9), implying that star formation has been going on beyond the main
sequence lifetime of OB stars. Population synthesis calculations using {\tt
STARBURST99} (Leitherer et al. 1999) confirm this suggestion. In Fig.
\ref{fig:sb99}, the dot-dashed line shows the spectral energy distribution
expected for a star cluster of total mass $5\times 10^6$ \msol, with a
Salpeter IMF predicted by {\tt STARBURST99}. Emission from such star
cluster fails to reproduce the observed luminosity of blob A.


\paragraph{Population  Synthesis}   We simulated blob A and blob B as  star forming
systems with SFR $\approx$ 1.22 \msol yr$^{-1}$ and SFR $\approx$ 0.42 \msol
yr$^{-1}$ respectively (estimated from L(\ha) and as reported in Table
\ref{tab:sfr}), using {\tt STARBURST99}. In Fig. \ref{fig:sb99}  we plot
the luminosity in the bands for which we have available  data (uncorrected
for internal extinction), along with the results of {\tt STARBURST99}
simulations (solid lines). A reasonable fit to the observed colors  and to
the absolute B magnitude of blob A is obtained for constant star formation
rate and an age of 7$\times 10^7$ yr.  This implies that the mass of the
blob is $M_{blob A} \ga 10^8$\msol\ (a lower limit since we do not know how
much of gas mass belongs to the blob). We are able to reproduce the total
blob A luminosity within a Starburst age significantly lower than the time
lapsed after A194S and N crossing. Not surprisingly, this is not possible
for A194S-A, the nucleus of A194S: if SFR$_{\rm A194S-A}\approx$1.55 \msol
yr$^{-1}$, after 7$\times 10^7$ yr a star forming system would be less
luminous than A194S by a factor $\approx$ 10 in the K band. This implies
that {\em the tidally stripped matter making blob A was, in origin,
predominantly gaseous}.

\subsubsection{The Supernova Rate From Radio Properties \label{rsn}:
Evidence of Substantial Obscuration}


Radio emission in star forming galaxies is expected from three major
sources: (1) radio supernov\ae\ (RSN\ae), (2) supernova remnants (SNRs),
and (3) cosmic ray electrons injected  through the Fermi acceleration
mechanism in the interstellar medium (see, e.g., Condon 1992 for an
excellent review). We can write the emitted power as a function of time as
$ P = \Sigma_{\rm k} \int_{t_{\rm min,k}}^{t_{\rm max,k}} dn(t)/dt~ P_{\rm
k}(t) dt$, where  the integration in summed over the three main radio
emission mechanisms, and $dn(t)/dt$\ is the number of supernov\ae\ per
year, and we assume that the age of the starburst is longer than the
maximum $t_{\rm max} \sim 10^7$ yr (see below).

A notable example of radio-emitting supernova has been  the RSN (with a type
Ibc progenitor) discovered in the circumnuclear starburst region of NGC
7469, a Seyfert 1 galaxy (Colina et al. 2001). The conditions of NGC 7469
may be ultimately similar to the ones of A194S-A. However, unless a
Starburst is extremely young, RSNae due to type II events should  dominate
the radio luminosity contribution due to SNae. We consider as
representative one of the best studied cases, SN 1979c in M100. The
luminosity is an order of magnitude less than RSN NGC 7469, $\nu P_\nu
\approx 5\times 10^{36}$\ergss, and the light curve can be modeled as flat
for a time $t_{\rm max,1} \approx$ 2 yr with a sharp rise and steep decline
(see also Yin \& Heeshen 1991; Mioduszewski et al. 2001). This seems to be,
at present, a reasonable albeit rough assumption. The contribution due to
SNR, $P_{\nu,\rm SNR}$\ is a strong function of time. The SNR diameter
depends      on time as $D = 4.3 \times 10^{-11} (E_0/n_{\rm
e,int})^\frac{1}{5} t^\frac{2}{5}$, where D is in pc and t in years (Clark
\& Caswell 1976; see e.g., Ulvestad 1982) up to a $t_{\rm max, 2}\approx
10^5$ yr. $E_0$\ is the total energy of a SN, and $n_{\rm e,int}$\ is the
electron density of the circumstellar medium ($n_{\rm e,int} \sim 1
cm^{-3}$). It is usually assumed that $E_0/n_{\rm e} \sim 10^{51}$ ergs
cm$^3$. We consider the observational surface brightness $\Sigma$ vs.  D
relationship in the form $\Sigma \approx 10^{-15}  D^{-3}$ W m$^{-2}$
Hz$^{-1}$ sr$^{-1}$\ (Ulvestad 1982, and references therein). Assuming a
power law index 0.8 typical of type II SNae  (Weiler et al. 1986; Colina et
al. 2001, and references therein), we obtain that the total power emitted
at 1.4 GHz by a SNR as a function of time is $ \nu P_{\rm \nu,1.4,SNR}
\approx 1.3 \times 10^{35} t^{-\frac{2}{5}}$   \ergss. A third term is due
to relativistic electrons injected by supernova shocks in the interstellar
medium. These ``cosmic ray" electrons are thought to  account for  $f$ =
90\%\ -- 94 \%\  of the total radio power output from a supernova remnant
(Condon 1992; Bressan et al. 2002), and  may radiate for $t_{\rm max,3}\la
10 ^7$ yr.


Assuming $dn(t)/dt$ independent on time, we obtain a rate $dn/dt\sim 2 \cdot
(1 - f) \sim 0.1-0.2 $ SN\ae\ yr$^{-1}$\ for the nuclear region, and $\sim 1
\cdot (1 - f) \approx 0.05 - 0.1$\ SN\ae\ yr$^{-1}$\ for blob A. The SN rate
deduced from radio power  is in any case much larger than the value
expected from stellar population synthesis, which predicts $dn(t)/dt
\approx 0.015$ SN\ae\ yr$^{-1}$\ for the nuclear regions of A194S, and
${dn(t)}/{dt} \approx 0.01$ SN\ae\ yr$^{-1}$\ for blob A. The SN rate
deduced from radio is $\sim 5-10$ times larger than the SN rate deduced
from optical properties. This hints at (1) the presence of an AGN, which
may significantly affect the total power of A194S; however, this does not
seem the case since the ratio between optical and radio SN rate on A194S-A
is similar to that of blob A;  at (2) obscuration; (3)  a short-lived
($\sim 10^7$ yr), post-Starburst phase in which the ratio SFR/$\nu P_\nu$\
reaches a minimum (Bressan et al. 2002). This is predicted for an impulsive
Starburst, and is due to the strong enhancement of the radio emission
following the Supernova explosion of the lowest-mass type-II supernova
progenitors (which are also the most frequent for standard IMFs).
Obscuration is most likely to play the major role. Especially in systems
like the nuclear regions of A194S, and blob A which are spheroidal in
appearance, \ha\ emission may be detected only from the star forming
regions nearer to be observer. The SFR measured from the FIR luminosity
is   $\ga$10 times larger than the SFR measured from \ha\ (Kennicutt 1998).
The  rate of SN\ae\ found from a search in the K band is also a factor
$\sim$ 5--10 then the rate estimated from  optical surveys (Maiolino et al.
2002; Mannucci et al. 2002, in preparation).


\section{Discussion \label{disc}}

\subsection{Cross-Fueling in Interpenetrating Encounters \label{int}}

Numerical simulations of head-on collisions between galaxies show that a
substantial mass of gas is ``splashed out" into a bridge connecting the two
 centers of potential. After the collision, the gas is re-accreted either from
the intruder and from the primary disk. The amount of material pushed out in
the collision depends on the relative orientation and impact parameter of
the encounter. The subsequent infall is spatially asymmetric and is
primarily located in a well defined streams. Most of the accreted gas ends
up in the central regions of the model galaxies (Struck 1997). However,
cross-fueling  is definitely not an easy  phenomenon to prove
observationally. In the framework of a galaxy pair, it implies  three
physical requirements: (1) that gas is stripped  from one galaxy; (2) that
the gas, stripped from one galaxy, is falling  toward the other; (3) that
the infalling gas is actually fueling a Starburst or an AGN.  These three
conditions may be met only in very special  systems, and may be
demonstrable  in many  fewer. At least, in the case of  collisional ring
galaxies systems  like Arp 194 it is easy to test whether cross  fueling is
indeed  occurring. First, the blob and the morphology of A194N provide
 evidence in favor of stripping.  The second condition is also
satisfied with reasonable certainty {\em  for blob A} (nothing can be said
on blob B and C). On blob A we are observing  \ha\ and \nii\ line
components whose redshift is increasing as the gas is approaching the
nucleus of A194S.  The third condition is also met, since the SFR$\approx$
4.5 \msol yr$^{-1}$ for A194S-A and $\sim 10$\msol  yr$^{-1}$ are typical
for estimates of SFR from \ha\ luminosity in Starburst galaxies.

The emission line feature observed between  blob A and A194S  in the
spectrum at P. A. =145$^\circ$\ is not unlike the one observed in NGC 7592,
in which the two interacting galaxies are much closer, and in which a bright
\ha\ filament connects the nuclei of the two components (Rafanelli \&
Marziani 1992; Marziani et al. 2001).  Interpenetrating  encounters  lead to
some easily demonstrable cases of tidal stripping. Al least two other cases
have been studied in detail, Kar 29 and ESO 253-IG026 (Marziani et al.
2001, Marziani et al. 1994). ESO 253-IG026 shows an impressive, bright
filament connecting the two galactic components.  Evidence of cross-fueling
has been collected  in other galactic pairs, especially in mixed
spiral/elliptical pairs in which a Starburst has been induced in the early
type component: for example, in AM 0327-285 (de Mello et al. 1995; 1996),
and in Arp 105 (Duc et al. 1997; see also Domingue 2001). We can study more
easily some systems containing a collisional ring galaxy since they are at a
special timing after contact and oriented favorably. Even if the conditions
are special, they provide  a laboratory to study accretion gas flows on 100
pc -- Kpc scales which cannot be obtained as easily in more complex systems
like most merging systems.


\subsection{Tidal Stripping, Tidal Dwarf Galaxies \& the Superwind Phenomenon \label{sw}}

Albeit the blob B and C look slightly fainter than A, it is conceivable
that they may be of comparable mass.  The total mass of the blobs can
therefore be  several 10$^8$ \msol. The ejection of stellar and gaseous
material into the intergalactic medium and its subsequent rearranging may
lead ultimately to the formation of self-gravitating tidal dwarf galaxies
(Duc et al. 2000). In this respect, it is interesting to note that the mass
of blob A is within the range of dwarf galaxies.

 The Arp 194 system  shows that a part of the orbital energy can be transferred to the gas motion
(e.g., Marziani et al.  1994 in the case of Kar 29,  Horellou \& Combes
2000) in the special case of a collisional ring galaxy. This is expected in
general from simulations of interacting galaxies. Orbital energy transfer
leads to vertical motions of gas from a system that, when unperturbed, was
dynamically very cold with a very low vertical velocity dispersion.
Filamentary structures emerging from  disk and dwarf galaxies are a
telltale  signature of  superwinds  (see for instance the spectacular case
of NGC 1808;  Phillips 1993).  Superwinds  are currently explained as due
to the kinetic  energy  injected by stellar winds and SN\ae\  around compact
Starburst regions (e.g. Heckman  et al. 1993; Heckman 2001).  This
explanation neglects the observational fact that the wide majority  of
superwind  galaxies  belong  to  strongly interacting systems (Marziani \&
Dultzin-Hacyan 2000).  A rather large fraction of superwind galaxies
(possibly 9 out of 20, of which 3 with obvious evidence of tidal stripping)
were found in merging systems. Small companions, leading to appreciable
perturbations, were also seen near $\approx \frac{1}{3}$ of the superwind
galaxies. Also in the case of minor mergers, the velocity dispersion in the
inner disk is expected to be significantly increased, and the disk
structure to become  unstable (Walker et al. 1996). Implications of
ongoing  interactions -- and of gravitationally-induced, non-rotational gas
motions --  have not been taken into account in the theory of galactic
superwinds. Even if it is possible to obtain a momentum and energy flow
consistent with the observed superwind properties from the mass  ejections
by stellar wind and SN\ae\ (Leitherer et al. 1992), several superwind
aspects -- including the possibility that the wind matter could escape from
the galactic potential well -- could  be influenced if  the out-of-disk
flow is eased by tidal forces, since most superwind flows are apparently
only ``marginally bound'' (Martin et al. 2000). There could be important
consequences  for the enrichment of any intra-cluster medium. We know the
relative  velocity between blob A and A194N-A.  We can estimate the kinetic
energy of tidally stripped matter as  $ E = \frac{1}{2}M_{\rm blob} \Delta
v^2 \approx 2.5\times 10^{54} M_8 \Delta v_{\rm 50}^2$ ergs. This is a
lower limit, since the blob mass could well be   much larger than
estimated. The kinetic energy imparted to blob A corresponds to the total
kinetic energy produced by $\sim 10^3-10^4$ SN\ae, and  it may be still a
small fraction of the total superwind energy ($\sim 10^{55} - 10^{57}$
ergs, which corresponds to the total output of 10$^4- 10^6$\ supernov\ae).
However, gravitational forces could have still an important impact. The
kinetic energy could easily be much higher, since $\Delta v$\ imparted to
the streaming gas could be much larger than the one assumed on the basis of
the blob A observations. The kinetic energy value for tidally stripped
matter and superwind outflow can therefore be comparable.  We have shown
that at least blob A and B can be thought as being entirely gaseous in
origin. The relevance of Arp 194 to the superwind phenomenon is that it
demonstrates observationally that gas motions can be perturbed to the point
of giving rise to a stream perpendicular to the galactic plane because of
gravitational acceleration.


There are other examples of a superwind-like outflow manifestly driven by
interactions. Kar 29 is likely to be the result of a high velocity
encounter (\dvr $\approx$ 1000 \kms) between an early-type galaxy (the
northern component) and a late-type spiral (see Hearn \& Lamb 2001 for a
review of observational data on this object). The stripped gas emerging
from the spiral southern component has not been captured by the elliptical,
and may be instead falling back toward the spiral as in a true ``galactic
fountain.'' Fraternali et al. (2001) found that the ``galactic fountain''
\hi\ motions in the spiral NGC 2403 are difficult to explain in the pure
superwind scheme without accounting for interactions. Also, van der Hulst
and Sancisi (1988) suggested that the high velocity \hi\ gas moving at high
speed perpendicular to the disk of Messier 101 could be explained in term
of a collision between a large gas mass and the disk of Messier 101 itself.
Similarly, some high velocity clouds observed in our Galaxy are likely due
to tidal effect related to the crossing of the Galaxy by the Large
Magellanic Cloud (e.g., Wakker \& van Woerden 1997). These findings can be
understood through the simulations mentioned in the introduction. We
conjecture that,  in mergers and in minor mergers, there could be an
important effect of gravitational forces on gas motion perpendicular to the
original disk plane.

\section{Conclusion}

We have analyzed in detail the photometric properties and the kinematics of
the strongly interacting system Arp 194. The main constituents of this
system are a collisional ring system (A194N) and an intruder (A194S) that
have experienced an interpenetrating, head-on encounter a few $ 10^8$ yr
ago. We have shown that tidally stripped gas is falling toward the center
of the intruder, A194S, and that it is fueling a strong nuclear and
circumnuclear Starburst. Arp 194 is therefore one of the few known objects
for which convincing evidence of cross-fueling exists. Considering that gas
is usually confined in a ``dynamically cold'' configuration in disk
galaxies, the Arp 194 case indicates transfer of orbital energy to the
internal motion of gas during the the encounter. We suggest that gas
motions in superwind galaxies -- which are mostly interacting systems --
could also be affected by the same mechanism.

Several aspects of the Arp 194 system deserve further scrutiny. The
morphology of A194N is fairly complex, and the distorted morphology of the
ring as well as the bright arc of A194N indicate the possibility of
perturbations by a third party. A full coverage with slit spectroscopy will
allow to confirm the presence of a second intruder galaxy located
approximately in correspondence of the northern side of the ring. High
spatial resolution spectroscopy may reveal more complex motions in
proximity of the nucleus of A194S. In the interacting galaxy pair NGC
1409/10, the ongoing mass transfer seems to follow a spiraling pattern
(Keel 2000; similar considerations may apply to NGC 7592: Hattori et al.
2002). There is no indication of an hidden AGN presence from our data, but
this result depends on the resolution of the FIRST survey data. It is
possible that higher resolution data may reveal  a high surface brightness
compact core. In addition, further radio observations could detect radio
supernova events: if $ dn/dt \approx 0.3$, and if a radio supernova event
remains detectable for $\approx$3 yr, we expect to reveal 1 event at any
observing time.

\acknowledgements P. M. acknowledges financial support from the Italian
MURST through  Cofin 00--02--004. D. D.- H. acknowledges grant IN 115599
PAPIIT-UNAM. We wish to thank Alessandro Bressan for  providing us with the
numbers of ionizing photons as a function of stellar spectral type and
Radoslav Zamanov for a fruitful discussion on pulsars.

\clearpage

\begin{deluxetable}{lcccc}
\tablewidth{0pt} \tablecaption{Log of SPM Observations \label{tab:obs}}
\tablehead{ \colhead{Date-Obs} & \colhead{U.T.\tablenotemark{a}} &
\colhead{E.T.\tablenotemark{b}} & \colhead{P.A.} & \colhead{Filter/Sp. Range} \\
& & \colhead{[s]}} \startdata
30-Jan-1995  & 11$^h$31$^m$  &     600  &  n. a. &  R            \\
30-Jan-1995  & 11$^h$47$^m$  &     600  &  n. a. &  R            \\
30-Jan-1995  & 12$^h$05$^m$  &     600  &  n. a. &  R            \\
30-Jan-1995  & 12$^h$19$^m$  &     600  &  n. a. &  R            \\
30-Jan-1995  & 12$^h$33$^m$  &     900  &  n. a. &  H$\alpha$+{\sc [Nii]}    \\
30-Jan-1995  & 12$^h$50$^m$  &     900  &  n. a. &  H$\alpha$+{\sc [Nii]}    \\
30-Jan-1995  & 13$^h$07$^m$  &     900  &  n. a. &  H$\alpha$+{\sc [Nii]}    \\
31-Jan-1995  & 09$^h$34$^m$  &     600  &  n. a. &  R            \\
31-Jan-1995  & 09$^h$46$^m$  &     900  &  n. a. &  B            \\
31-Jan-1995  & 10$^h$03$^m$  &     900  &  n. a. &  B            \\
31-Jan-1995  & 10$^h$27$^m$  &     900  &  n. a. &  B            \\
2-Feb-1995   & 11$^h$54$^m$  &     1800 &   145$^\circ$  & 6165--7246 \AA \\
2-Feb-1995   & 12$^h$25$^m$  &     1800 &   145$^\circ$  & 6165--7246 \AA \\
2-Feb-1995   & 13$^h$04$^m$  &     1800 &   118$^\circ$  & 6165--7246 \AA \\
2-Feb-1995   & 13$^h$32$^m$  &     1800 &   118$^\circ$  & 6165--7246 \AA \\
\enddata
\tablewidth{\textwidth} 
\tablenotetext{a}{Universal Time at exposure start.}
\tablenotetext{b}{Exposure Time duration in seconds for each single frame.}

\end{deluxetable}

\begin{deluxetable}{llccccccccccc}\label{tab:mf}
\tabletypesize{\scriptsize} \tablewidth{0pt} \tablecaption{Multifrequency
Data for the Arp 194 System } \tablehead{ \colhead{Galaxy} &
\colhead{Region Id.} & \colhead{\ha+[N{\sc ii}]\tablenotemark{a}} &
\colhead{B} & \colhead{R} &\colhead{B--R} & \colhead{J\tablenotemark{b}} &
\colhead{K\tablenotemark{b}} & \colhead{f$_\nu$(1.4 GHz)} \\
& & \colhead{[ergs s$^{-1}$ cm$^{-2}$]} & & & & & & [mJy] } \startdata
A194S & Total \tablenotemark{c}          &  2.29E-13 & 15.17 & 14.39 &    0.78 & 14.22    &        12.67& $\ga$5.35 \\
A194S & Nucleus\tablenotemark{e}     &  7.94E-14 & 16.31 & 15.84 &    0.47 & 15.6     &        13.94& 4.4\tablenotemark{d} \\
Arp 194  & Blob A\tablenotemark{e}      &  6.26E-14 & 18.34 & 17.87 &    0.47 & $>$18.35\tablenotemark{f}  &     $>$16.35\tablenotemark{f} & 2.79 \\
Arp 194  & Blob B\tablenotemark{e}      &  2.17E-14 & 18.18 & 18.06 &    0.12 & $>$18.35\tablenotemark{f}  &     $>$16.35\tablenotemark{f} & \nodata \\
Arp 194  & Blob C                              &  1.41E-14 & 17.98 & 17.72 &    0.26 & $>$18.35\tablenotemark{f}  &     $>$16.35\tablenotemark{f} & \nodata \\
\\
A194N & Total                               &  1.38E-13 &  14.37 & 13.58 &   0.79 &  13.88   &   12.11    & $>$0.40       \\
A194N & A194N-A ``Nucleus''\tablenotemark{e}  &  $\la$2.13E-14\tablenotemark{g} &  17.11 & 16.05 &   1.06 &  15.94   &   14.26    &    0.40   \\
A194N & A194N-B  &  1.38E-14 & 17.51 &  16.34  & 1.17  &  15.51   &   13.62 & \nodata  \\
A194N & Blob D       &  \nodata                          & 18.04  & 17.36   & 0.68 &   \nodata &   \nodata   & \nodata  \\
A194N & Arc Spot A  &  1.10E-14 & 18.25 &   17.87   & 0.38 & \nodata   & \nodata   & \nodata  \\
A194N & Arc Spot B  &  7.14E-15 & 17.96 &   17.38   & 0.58 & \nodata   & \nodata   & \nodata  \\
A194N & Arc Spot C  &  9.61E-15 & 18.03 &   17.15   & 0.88 & \nodata   & \nodata   & \nodata  \\

\enddata

\tablewidth{\textwidth} \tablenotetext{a}{The [Nii]\l 6548 line falls right
at the border of the employed narrow band filter; however considering that
I([Nii]\l 6583)/I(\ha)$\approx$0.4 almost everywhere, this implies that the
\ha\ + \nii\ fluxes should be corrected by $\la$ 10\%\ (below our estimated
uncertainties even for the brightest \ha\ emitting regions). No correction
was applied to the values reported in this Table.} \tablenotetext{b}{Near
Infrared photometry by Bushouse \& Stanford (1992).}
\tablenotetext{c}{Excludings blobs.} \tablenotetext{d}{Value measured on
FIRST with an aperture  of 5".4.} \tablenotetext{e}{ Emitting Region covered
fully or in part with long slit spectroscopy (see Table \ref{tab:spec}).}
\tablenotetext{f}{Regions not visible in the isophotal contour map by
\cite{bushouse}. The lower limit to the magnitude has been estimated from
the lowest level isophote plotted by \cite{bushouse}, which are $\sigma
\approx$ 20.5 mag arcsec$^2$ and 18.5 mag arcsec$^2$ for J and K band
respectively. With a scale 1.35"/pixel, the magnitude of a very faint
object contoured in a square of 1 pixel of radius would lead to   $ m
\approx$ --2.15 + $\sigma$. } \tablenotetext{g}{\ha\ measured on a larger
area than B and R.}
\end{deluxetable}

\begin{deluxetable}{llcccccccccccccccccccccc}
\rotate \tabletypesize{\scriptsize} \tablewidth{780pt}
\tablecaption{Emission Line Regions in   Arp 194S \label{tab:spec}}
\tablehead{ \colhead{Galaxy} & \colhead{Region Id.} & \colhead{P. A.}&
\colhead{\vr$_{\rm Hel}$} & \multicolumn{3}{c}{[\sc{Nii}]\l 6548} &&
\multicolumn{3}{c}{\ha} && \multicolumn{3}{c}{[\sc{Nii}]\l 6583} &&
\multicolumn{3}{c}{[\sc{Sii}]\l 6716} && \multicolumn{3}{c}{[\sc{Sii}]\l
6731} \\ \cline{5-7} \cline{9-11} \cline{13-15} \cline{17-19} \cline{21-23}
&& ($^\circ$)& [\kms] & \colhead{Flux\tablenotemark{a}} &
\colhead{W\tablenotemark{b}} & \colhead{FWHM\tablenotemark{c}} &&
\colhead{Flux\tablenotemark{a}} & \colhead{W\tablenotemark{b}} &
\colhead{FWHM\tablenotemark{c}} && \colhead{Flux\tablenotemark{a}} &
\colhead{W\tablenotemark{b}} & \colhead{FWHM\tablenotemark{c}} &&
\colhead{Flux\tablenotemark{a}} & \colhead{W\tablenotemark{b}} &
\colhead{FWHM\tablenotemark{c}} && \colhead{Flux\tablenotemark{a}} &
\colhead{W\tablenotemark{b}} & \colhead{FWHM\tablenotemark{c}} } \startdata
A194S & A  Nucleus\tablenotemark{d} & 145& 10502$\pm$3&  8.4 & 24 & 170 && 64.5 & 170 &170 && 26 & 68.5 & 170 && 11.9 & 33.6 & 220 && 8.4 &24 & 220 \\
A194  & Blob A\tablenotemark{e} & 145& 10532$\pm$7 &1.2 & \nodata & 140 && 15.6 & \nodata & $\la$110 && 3.61 & \nodata & 110 && 2.1\tablenotemark{f} & \nodata& 200: && 1.6 & \nodata & 110 \\
A194N & A ``Nucleus''\tablenotemark{g} &145& 10435:\tablenotemark{h}& \nodata & \nodata & \nodata && 0.46 & 5 & 250 && 0.27 & 3 & 250 &&  \nodata & \nodata & \nodata &&  \nodata & \nodata & \nodata\\
A194N & A ``Nucleus''\tablenotemark{g}& 118&   10430$\pm$20&   \nodata & \nodata & \nodata && 0.84 & 8.3 & 160 && 0.6 & 6.6 & 220&& \nodata & \nodata & \nodata && \nodata & \nodata & \nodata   \\
\enddata
\tablewidth{\textwidth}

\tablenotetext{a}{Observed  flux in units of 10$^{-15}$  ergs s$^{-1}$
cm$^{-2}$, uncorrected   for  redshift  and  internal   extinction. Note
that the correspondence between the long-slit emitting regions and the
regions measured on the narrow and broad band images labeled with the same
name is rather rough (it can be checked in Fig. \ref{fig:ident}\ where the
slit position is marked).} \tablenotetext{b}{Equivalent width in \AA.}
\tablenotetext{c}{Full Width at Half Maximum in \kms\ corrected for
instrumental  broadening $\approx$ 2.5 \AA\  FWHM.} \tablenotetext{d}{Only
on this region a test on the accuracy of the spectra flux calibration can
be done comparing the flux of the nucleus of A194S
 with the \ha\ flux measured in the narrow
band image (Tab. \ref{tab:mf}), since the summation was done on an area of
similar value value ($\approx$ 25 arcsec$^2$). The agreement
 is very good even if the apertures are of different shape.}

\tablenotetext{e}{Almost  no continuum; equivalent width not defined. }

\tablenotetext{f}{Flux estimated from peak intensity ratio since the line
profile is contaminated by a blemish.}

\tablenotetext{g}{Only partially in the slit.}

\tablenotetext{h}{Double-peaked \ha\ and \nii\ lines. A second component,
partially resolved, peaks at \vr$_{\rm Hel} \approx$ 10566 \kms.}
\end{deluxetable}

\begin{deluxetable}{llccccccccc}
\tabletypesize{\scriptsize} \tablewidth{0pt} \tablecaption{Star Formation
Results  for the Arp 194 System \label{tab:sfr}} \tablehead{
\colhead{Galaxy} & \colhead{Region Id.} & \colhead{$M_B$\tablenotemark{a}} &
\colhead{L(\ha)} & \colhead{SFR\tablenotemark{b}}
&\colhead{Q(H)\tablenotemark{c}} & \colhead{N(OB)\tablenotemark{d}} &
\colhead{${\cal{M}}_{\rm tot}$\tablenotemark{e}}\\
& & & \colhead{[\ergss]} & \colhead{[\msol yr$^{-1}$]} & \colhead{[s$^{-1}$]} & & \colhead{[\msol]}\\
}  \startdata
A194S & Total                               & $-$21.05  &  5.7$\times 10^{41}$  & 4.5   &  4.8$\times 10^{53}$    &  100000     & 1.8$\times 10^{7}$\\ %
A194S & A Nucleus                           & $-$19.91  &  2.0$\times 10^{41}$  & 1.6  &  1.7$\times 10^{53}$    &  34000      & 6.3$\times 10^{6}$\\%
A194  & Blob A                              & $-$17.88  &  1.6$\times 10^{41}$  & 1.2  &  1.3$\times 10^{53}$    &  27000      & 5.0$\times 10^{6}$\\
A194  & Blob B                              & $-$18.04  &  5.4$\times 10^{40}$   & 0.4  &  4.6$\times 10^{52}$    &  9000      & 1.7$\times 10^{6}$\\
A194  & Blob  C                             & $-$18.24  &  3.5$\times 10^{40}$ & 0.3    &  3.0$\times 10^{52}$    &  6000      & 1.1$\times 10^{6}$ \\
A194N & Total                               & $-$21.85  &  3.4$\times 10^{41}$  & 2.7   &  2.9$\times 10^{53}$    &  60000     & 1.1$\times 10^{7}$\\
A194N & A ``Nucleus''                       & $-$19.11  &  5.3$\times 10^{40}$  & 0.4    &  4.5$\times 10^{52}$    &  9000      & 1.7$\times 10^{6}$\\
A194N & B                                   & $-$18.19  &  3.4$\times 10^{40}$  & 0.3   &  2.9$\times 10^{52}$    &  6000      & 1.1$\times 10^{6}$\\
A194N & Blob D                              & $-$18.18 &    \nodata\tablenotemark{f}              & \nodata  &  \nodata               &  \nodata   & \nodata                                                      \\
A194N & Arc Spot A                          & $-$17.97  &  2.7$\times 10^{40}$  & 0.2   &  2.3$\times 10^{52}$    &  5000      & 8.7$\times 10^{5}$\\
A194N & Arc Spot B                          & $-$18.26  &  1.8$\times 10^{40}$  & 0.15   &  1.5$\times 10^{52}$    & 3000       & 5.7$\times 10^{5}$\\
A194N & Arc Spot C                          & $-$18.19  &  2.4$\times 10^{40}$  & 0.2    &  2.0$\times 10^{52}$    & 4000      & 7.63$\times 10^{5}$\\
\enddata
\tablewidth{\textwidth}

\tablenotetext{a}{Blue absolute  magnitude $M_B$.} \tablenotetext{b}{Total
SFR from 0.1 to 100 \msol\ computed for a Salpeter IMF from L(\ha): SFR
$\approx 7.9  \times 10^{-42}$ L(\ha) \msol yr$^{-1}$ \ (Kennicutt 1998).}
 \tablenotetext{c}{The
number of ionizing photons Q(H) has been computed from L(\ha) following case
B of nebular theory, assuming T$_e \approx$10000$^\circ$K, and no photon
escaping the nebula.} \tablenotetext{d}{Number of OB stars, i.e., stars in
the mass range 10 \msol -- 100 \msol, computed from Q(H) assuming a
relationship between each star Q(H) and mass from Kurucz's models (A.
Bressan, private communication), and a Salpeter IMF.}

\tablenotetext{e}{Total mass of IMF stars from 0.1 \msol to 100 \msol\
assuming a Salpeter IMF.}

\tablenotetext{f}{Very faint \ha\ emission.}
\end{deluxetable}

\clearpage


\end{document}